\documentclass[epj,referee]{svjour}
\usepackage{epsfig,psfrag}
\begin{document}
%
\title{Exactly solvable toy model for the pseudogap state}
\author{Lorenz Bartosch and Peter Kopietz} 
\institute{
Institut f\"{u}r Theoretische Physik, Universit\"{a}t G\"{o}ttingen,
Bunsenstrasse 9, 37073 G\"{o}ttingen, Germany}
\date{March 15, 2000}
\abstract{
We present an
exactly solvable toy model which describes the emergence of a pseudogap
in an electronic system
due to a fluctuating off-diagonal order parameter.
In one dimension our model reduces to the
fluctuating gap model (FGM)
with a gap $\Delta ( x )$ that is constrained to be of the form
$\Delta ( x ) = A e^{i Q x}$,
where $A$ and $Q$ are {\it{random variables}}. 
The FGM was 
introduced by Lee, Rice and 
Anderson [Phys. Rev. Lett. {\bf{31}}, 462 (1973)]
to study fluctuation effects in Peierls chains.
We show that their perturbative results  for the average
density of states are exact for our toy model if we assume
a Lorentzian  probability distribution for $Q$ and ignore
amplitude fluctuations.
More generally, choosing the probability distributions of
$A$ and $Q$ such 
that the average of $ \Delta (x )$ vanishes and its covariance is
$\langle  {\Delta} ( x ) {\Delta}^{\ast} ( x^{\prime} ) \rangle
= \Delta_s^2 \exp[ { - | x - x^{\prime} | / \xi }]$,
we study the combined effect of phase and amplitude fluctutations
on the low-energy properties of Peierls chains.
We explicitly calculate the average density of states, the 
localization length, the average single-particle Green's function, and
the real part of the average conductivity.
In our model phase fluctuations generate delocalized 
states at the Fermi energy, which give rise to a
finite Drude peak in the conductivity.
We also find that the interplay between phase and amplitude
fluctuations leads to a weak logarithmic singulatity
in the single-particle spectral function at the bare quasi-particle energies.
In  higher dimensions our model
might be relevant to describe the 
pseudogap state in the underdoped cuprate superconductors.}
\maketitle
%
%
%
%
%

\section{Introduction}
\label{sec:intro}

The physical origin of the pseudogap behavior observed
in the normal state of the high-temperature cuprates is
still controversial. Several mechanisms have been
proposed. According to Schmalian et al. \cite{Schmalian98}
the normal state of the underdoped cuprates
can be modeled by a nearly antiferromagntic Fermi liquid, and
the experimentally observed pseudogap behavior is closely
related to strong antiferromagnetic spin fluctuations.
An alternative explanation, which has been advanced by
Emery and Kivelson \cite{Emery95} relates the pseudogap behavior
to precursor superconducting fluctuations.
In this scenario thermal
fluctuations of the phase of the superconducting
order parameter are responsible for a destruction
of superconductivity above the transition temperature
$T_c$. However, in a wide range of temperatures $T > T_c$
the local amplitude of the superconducting
gap is finite. In this paper we shall propose a simple
exactly solvable phenomenological model which
describes the destruction of phase coherence due
to phase and amplitude fluctuations of the superconducting order
parameter in the pseudogap state.

To study superconducting fluctuations in a normal metal
one can start with the Gorkov equation
for the $2 \times 2$ matrix Green's function for
electrons with energy dispersion $\epsilon ( {\bf{k}} )$
that are coupled to a space-dependent  complex pairing field
$\Delta ( {\bf{r}} )$ \cite{Abrikosov63},
 \begin{equation}
 [ \omega - \hat{H}_{\bf{r}} ] \,
 {\cal{G}}^{(d=3)} ( {\bf{r}}, {\bf{r}}^{\prime}, \omega )
 = \delta ( {\bf{r}} - {\bf{r}}^{\prime} ) \sigma_0
 \label{eq:Gorkov}
 \; ,
 \end{equation}
 \begin{equation}
\hat{H}_{\bf{r}} =
 \left( \begin{array}{cc}
  \epsilon ( - i \nabla_{\bf{r}} ) - \mu  & \Delta ( {\bf{r}} ) \\
  \Delta^{\ast} ( {\bf{r}} ) &  \epsilon (  i \nabla_{\bf{r}} )
  - \mu
  \end{array} \right)
 \label{eq:HamiltonianGorkov}
 \; .
 \end{equation}
Here, $\sigma_0$ is the $2 \times 2$ unit matrix and
$\mu $ is the chemical potential.
In the absence of true superconducting
long-range order the pairing field
 $\Delta ({\bf{r}})$ can be considered as a random variable with
zero average and correlations that fall off exponentially with distance,
  \begin{equation}
 \langle \Delta ( {\bf{r}} ) \rangle   = 0
 \; ,
 \label{eq:deltaavG}
 \end{equation}
 \begin{eqnarray}
 \langle \Delta ( {\bf{r}} ) \Delta^{\ast} ( {\bf{r}}^{\prime} ) \rangle
 & \equiv &
 \frac{ \int {\cal{D}} \{ \Delta \} e^{- S \{ \Delta \} } \Delta ( {\bf{r}} ) 
 \Delta^{\ast} ( {\bf{r}}^{\prime} ) }{
 \int {\cal{D}} \{\Delta \} e^{- S \{ \Delta \} } } 
 \nonumber
 \\
 & = &
 \Delta_s^2 e^{ - | {\bf{r}} -  {\bf{r}}^{\prime} | / \xi }
 \; .
 \label{eq:deltacovG}
 \end{eqnarray}
Here, $S \{ \Delta \}$ is the Ginzburg-Landau functional of the
order parameter field, $\xi$ is the correlation length,
and the energy scale $\Delta_s$ characterizes the strength of the
correlations.

To simplify the algebra and to make contact with
other theoretical work on pseudogap physics, we shall focus in this
work on the semiclassical limit
of the Gorkov equation, which are related to the so-called Andreev
equation \cite{Andreev64}. In the weak coupling limit, where
$| \Delta ({\bf{r}})|$ is small
compared with the chemical potential, we may 
linearize the energy dispersion in Eq.\ (\ref{eq:Gorkov}) for
wave-vectors ${\bf{k}}$ close to the Fermi surface,
provided we are only interested
in long-wavelength, low-energy properties of the system.
In the semiclassical limit it is useful to decompose the position
vector as  
${\bf{r}} = x {\bf{n}} + {\bf{r}}_{\bot}$
where ${\bf{n}}$ is a unit vector in the direction of the momentum
of the electron, and ${\bf{r}}_{\bot}$ is orthogonal to ${\bf{n}}$.
Writing  $\partial_x = {\bf{n}} \cdot \nabla_{\bf{r}}$,
Eqs.\ (\ref{eq:Gorkov}) and (\ref{eq:HamiltonianGorkov}) can be replaced
by an effective one-dimensional problem \cite{Andreev64}
 \begin{equation}
 [ \omega - \hat{H}_x ] \,
 {\cal{G}} ( x, x^{\prime}, \omega )
 = \delta ( x -x^{\prime} ) \sigma_0
 \label{eq:Andreev}
 \; ,
 \end{equation}
 \begin{equation}
 \hat{H}_x   =    
 \left( \begin{array}{cc}
  - i v_F \partial_x  & \Delta ( x )  \\
  \Delta^{\ast} ( x ) & i v_F \partial_x 
  \end{array} \right)
 \label{eq:Hamiltonian}
 \; .
 \end{equation}
We shall refer to Eq.\ (\ref{eq:Hamiltonian})
as the Hamiltonian of the fluctuating gap  model
(FGM).
All quantities depend now
parametrically on ${\bf{r}}_{\bot}$ and ${\bf{n}}$. 
Physical observables should be averaged over all directions
of ${\bf{n}}$. In this paper we  shall only consider
the effective one-dimensional problem defined 
by Eqs.\ (\ref{eq:Andreev}) and (\ref{eq:Hamiltonian}).
We require that the first and the second
moments
of the fluctuating gap $\Delta (x )$ are given by
  \begin{equation}
 \langle \Delta ( x ) \rangle   = 0
 \; ,
 \label{eq:deltaav}
 \end{equation}
 \begin{equation}
 \langle \Delta ( x ) \Delta^{\ast} ( x' ) \rangle
 = \Delta_s^2 e^{ - | x -  x^{\prime} | / \xi }
 \; .
 \label{eq:deltacov}
 \end{equation}
In the following, we shall construct a special  non-Gaussian
probability distribution of $\Delta ( x )$ 
satisfying Eqs.\ (\ref{eq:deltaav}) and (\ref{eq:deltacov})
for which Eq.\ (\ref{eq:Andreev}) 
can be solved exactly.
Moreover, as will be briefly discussed in 
Sec.\ \ref{sec:conclusions}, it is straightforward
to generalize our model to
dimensions $d >1$ and to arbitrary energy dispersions
$\epsilon ( {\bf{k}})$, although
the calculation of physical quantities becomes more tedious.

Apart from its relevance in the semiclassical theory of
superconductivity, the
problem defined by Eqs.\ (\ref{eq:Andreev}) to (\ref{eq:deltacov})
describes also the low-energy physics in 
quasi-one-dimensional Peierls and spin-Peierls systems \cite{Lee73,Bunder99}. 
Lee, Rice and Anderson \cite{Lee73} used this model to study fluctuation
effects close to the Peierls transition. In this case 
$\Delta ( x )$ can be identified with the fluctuating Peierls order
parameter, and the two diagonal elements
in our Hamiltonian (\ref{eq:Hamiltonian}) 
represent the kinetic energy of the electrons in the vicinity
of the two Fermi points $\pm k_F$.
Physical quantities should again be averaged
over the probability distribution
of $\Delta (x )$, which can be obtained from the Ginzburg-Landau expansion \cite{Lee73}.
Within the Gaussian approximation, the truncated Ginzburg-Landau functional 
in the disordered phase is of the form
 \begin{equation}
 S \{ \Delta \} =  
 \int \frac{d q}{ 2 \pi} \,
 \frac{ 1 + q^2 \xi^2 }{ 2 \Delta_s^2 \xi} \, \Delta_q^{\ast} \Delta_q 
 \; ,
 \label{eq:SGL}
 \end{equation}
where
 \begin{equation}
 \Delta_q = \int dx \, e^{- i q x }  \Delta (x )
 \;  .
 \label{eq:deltaqdef}
 \end{equation}
One easily verifies that
Eqs.\ (\ref{eq:deltaav}) and (\ref{eq:deltacov}) are indeed satisfied. 
Note that for commensurate Peierls chains the order parameter field
can be chosen real, while it is complex for
incommensurate chains. In this work we
shall focus on the incommensurate case, where 
zero-energy states and the associated Dyson singularities
are absent \cite{Bartosch99a,Bartosch99b}.
Lee, Rice and Anderson  treated the effect of the order parameter
fluctuations on the average electronic density
of states (DOS) $\langle \rho ( \omega ) \rangle$ 
within the Born approximation. Within this approximation
one finds that, in the regime where the dimensionless parameter
 \begin{equation}
 \bar{\gamma} \equiv \frac{v_F}{2 \Delta_s \xi}
 \label{eq:gammabardef}
 \end{equation}
is small compared with unity,
the DOS develops a pseudogap for
$ | \omega |
{ \raisebox{-0.5ex}{$\; \stackrel{<}{\sim} \;$}}  \Delta_s$,
with a  minimum given by \cite{Chandra89}
 \begin{equation}
 \frac{\langle \rho ( 0 ) \rangle^{\rm pert}}{  \rho_0}
   =  \frac{ \bar{\gamma}}{ \sqrt{1 + \bar{\gamma}^2}} 
 \; .
 \label{eq:rhoBorn}
 \end{equation}
Here,
 \begin{equation} 
 \rho_0 = \frac{1}{ \pi v_F }
 \end{equation}
is the DOS for $\Delta (x) = 0$, which is a constant due to the
linearization of the energy dispersion. Note that
Eq.\ (\ref{eq:rhoBorn}) predicts for $\bar{\gamma} \ll 1$
to leading order
 \begin{equation}
 \frac{\langle \rho ( 0 ) \rangle^{\rm pert}}{  \rho_0}
   \sim  \bar{\gamma} \propto \xi^{-1} 
 \; ,
 \label{eq:rhoBornsmall}
 \end{equation}
which disagrees with a 
non-perturbative result by Sadovskii \cite{Sadovskii79},
who found for the model
defined by Eqs.\ (\ref{eq:Andreev}) to (\ref{eq:deltacov})
for a Gaussian distribution of $\Delta (x )$
 \begin{equation}
 \frac{ \langle \rho ( 0 ) \rangle^{\rm Sadovskii} }{ \rho_0}  \approx
 0.541 \times  [ {2 \bar{\gamma}} ]^{1/2} \propto \xi^{-1/2}
 \; .
 \label{eq:rhoSadovskii}
 \end{equation}
However, the algorithm constructed
by Sadovskii \cite{Sadovskii79} is not  exact \cite{Tchernyshyov99,Bartosch99a}, 
so that it is not clear whether Eq.\ (\ref{eq:rhoSadovskii})
is correct or not. To clarify this point, we have
recently developed an exact numerical algorithm for
calculating
the DOS of the FGM \cite{Bartosch99b}. For a Gaussian distribution
of $\Delta (x)$ with zero average and covariance given by
Eq.\ (\ref{eq:deltacov}) the result is
 \begin{equation}
 \frac{ \langle \rho ( 0 ) \rangle^{\rm Gauss} }{ \rho_0}  \approx
 a  [ 2 \bar{\gamma} ]^{b} \propto \xi^{-b}
 \; ,
 \label{eq:rhoGauss}
 \end{equation}
where
 \begin{equation}
 a  =  0.6397 \pm 0.0066
 \; \; , \; \; 
 b  =  0.6397 \pm 0.0024
 \label{eq:bdef}
 \; .
 \end{equation} 
Hence, for Gaussian disorder with a finite correlation
length both
perturbation theory and Sadovskii's algorithm do not give the
correct $\xi$-dependence of the average DOS at the Fermi energy.
Another attempt to investigate the discrepancy between
Eqs.\ (\ref{eq:rhoBorn}) and (\ref{eq:rhoSadovskii}) numerically was 
recently made by Millis and Monien \cite{Millis99}.
They found  for the exponent $b$ in Eq.\ (\ref{eq:rhoGauss})
a value between $2/3$ and $1$,
which is outside our error-bars in Eq.\ (\ref{eq:bdef}).
Note, however,  that Millis and Monien studied a lattice regularization of the
continuum  model (\ref{eq:Hamiltonian}), and no attempt was made to
carefully relate the bare parameters that appear in the lattice and
the continuum models.
In this  work we shall show 
that the exponent characterizing 
the behavior of the DOS at the Fermi energy on 
$\xi$ is non-universal in the sense
that it depends on the precise form of the probability distribution
of the fluctuating gap. In particular, the non-Gaussian terms
in the Ginzburg-Landau functional can change
the numerical value of this exponent, so that the
behavior given in Eqs.\ (\ref{eq:rhoGauss}) and (\ref{eq:bdef}) 
can only be expected to be correct for Gaussian disorder.

Finally, it should be mentioned that a generalization of the model 
defined in Eqs.\ (\ref{eq:Andreev}) to (\ref{eq:deltacov}) 
has been used in Ref. \cite{Schmalian98} to explain the pseudogap
behavior in the cuprates within antiferromagntic Fermi liquid theory.  
Then the scalar field
$\Delta (x )$ should be replaced by a matrix field
$\sum_i {{S}}_i ( x )  {\sigma}_i $,
where ${\sigma}_i $ are the 
Pauli matrices, and the fields ${{S}}_i ( x)$  
represent the components of the  antiferromagnetic spin density field.
In fact, the recent interest in the non-perturbative approach invented
many years ago 
by Sadovskii \cite{Sadovskii79} is  motivated
by its possible relevance to the cuprate superconductors.

\section{Exact Green's function of the fluctuating gap model 
for $\Delta (x ) = A e^{ i Q x}$}

In this section we shall solve Eq.\ (\ref{eq:Andreev}) exactly
for a special form of the probability distribution
of $\Delta ( x )$ which is constructed such
that its covariance is given by Eq.\ (\ref{eq:deltacov}).
To begin with, let us perform the following gauge 
transformation \cite{Brazovskii76},
 \begin{equation}
 {\cal{G}} ( x , x^{\prime} , \omega )
 = e^{\frac{i}{2} \alpha ( x ) \sigma_3 }
 \tilde{\cal{G}} ( x , x^{\prime} , \omega )
  e^{- \frac{i}{2} \alpha ( x^{\prime} ) \sigma_3 }
  \; ,
  \label{eq:gaugetrafo}
  \end{equation}
where the gauge function $\alpha ( x )$ will be specified shortly.
From Eq.\ (\ref{eq:Andreev}) we find that the
transformed Green's function
$ \tilde{\cal{G}} ( x , x^{\prime} , \omega )$ satisfies
 \begin{eqnarray}
 \Big[ 
  \omega -  \frac{v_F}{2} \frac{ d \alpha ( x )}{dx}  + i v_F \partial_x \sigma_3  
  -  \Delta (x ) e^{-  i \alpha ( x ) } \sigma_{+}
 \Big.
 \nonumber
 \\
 &  & \hspace{-67mm} 
 \Big.
  {} - \Delta^{\ast} ( x ) e^{  i \alpha ( x ) } \sigma_{-}
  \Big] 
 \tilde{{\cal{G}}} ( x , x^{\prime} , \omega )
 = \delta ( x - x^{\prime} ) \sigma_0
 \; .
 \label{eq:G1def}
 \end{eqnarray}
Suppose now that $\Delta ( x )$ is of the form
 \begin{equation}
 \Delta ( x ) = A e^{ i  Q x  }
 \; ,
 \label{eq:deltasimple}
 \end{equation}
where $A$ and $Q$ are both random but  
independent of $x$.
Then the $x$-dependence of
$\Delta (x )$ in Eq.\ (\ref{eq:G1def}) can be removed by choosing
$\alpha ( x ) = Q x$.
Moreover, with this choice
the second term on the left-hand side of 
Eq.\ (\ref{eq:G1def}) reduces to a constant
 \begin{equation}
 \frac{v_F}{2} \frac{ d \alpha (x)}{dx} =
 \frac{v_F Q}{2} \equiv \eta
 \label{eq:etadef}
 \; ,
 \end{equation}
so that
 \begin{eqnarray}
 \left[ 
  \omega -  \eta  + i v_F \partial_x \sigma_3  
  -  A \sigma_{+}
  - A^{\ast}  \sigma_{-}
  \right] 
 \tilde{\cal{G}} ( x , x^{\prime} , \omega )
 & & 
 \nonumber
 \\
 & & \hspace{-20mm}
 = \delta ( x - x^{\prime} ) \sigma_0
 \; .
 \label{eq:G1def2}
 \end{eqnarray}
Thus, a phase of the order-parameter varying linearly in space can be
absorbed by a finite shift of the energy.
Eq.\ (\ref{eq:G1def2}) is translational invariant and is easily solved 
by a Fourier transformation,
 \begin{equation}
 \tilde{{\cal{G}}} ( x , x^{\prime} , \omega )
  = \int \frac{ d q}{2 \pi} e^{i q ( x - x^{\prime} ) }
 \tilde{\cal{G}} ( q , \omega )
 \; ,
 \label{eq:G1FT}
 \end{equation}
 \begin{eqnarray}
 \tilde{\cal{G}} ( q , \omega ) & = &
 \frac{1}{  ( \omega - \eta )^2 - (v_F q )^2 -  | A |^2 }
 \nonumber
 \\
 & & \times
 \left( \begin{array}{cc}
 \omega - \eta + v_F q & A \\
 A^{\ast} & \omega - \eta - v_F q 
 \end{array}
 \right)
 \; .
 \label{eq:G1res}
 \end{eqnarray}
Combining
Eqs.\ (\ref{eq:gaugetrafo}), (\ref{eq:G1FT}) and (\ref{eq:G1res})
and defining
 \begin{equation}
 {\cal{G}} ( q , q^{\prime} , \omega )
 = \int dx \int d x^{\prime} e^{- i ( q x - q^{\prime} x^{\prime} ) }
 {\cal{G}} ( x , x^{\prime} , \omega )
 \; ,
 \label{eq:Gft}
 \end{equation}
we finally obtain
 \begin{eqnarray}
 {\cal{G}} ( q , q^{\prime} , \omega )
  & = &  
 \left(
 \begin{array}{cc}
 {\displaystyle
 \frac{ 
  2 \pi \delta ( q - q^{\prime} ) 
 [ \omega - 2 \eta + v_F q  ]}{ 
 [ \omega -2 \eta + v_F q  ][ \omega - v_F q ] - | A |^2 } }
 & \hspace{2mm}
 {\displaystyle
 \frac{ 2 \pi \delta ( q - q^{\prime} - Q )  A}{ 
 [ \omega - 2 \eta + v_F  q  ][ \omega - v_F q ] - | A |^2} }
 \\
 {\displaystyle \rule [0mm]{0mm}{8mm} 
 \frac{  2 \pi \delta ( q - q^{\prime} + Q ) 
 A^{\ast}}{ 
 [ \omega - 2 \eta - v_F  q  ][ \omega + v_F q ] - | A |^2} }
 & \hspace{2mm}
 {\displaystyle
 \frac{  2 \pi \delta ( q - q^{\prime} ) 
 [ \omega -2 \eta - v_F  q ] }{ 
 [ \omega -2 \eta - v_F  q ][ \omega + v_F q ] - | A |^2}
 } 
 \end{array}
 \right)
 \; .
 \nonumber
 \\
 & &
 \label{eq:Gqqres}
 \end{eqnarray}
The crucial observation is now that,
in spite of the simple form  (\ref{eq:deltasimple})
of $\Delta ( x )$, it is still possible to
satisfy
Eqs.\ (\ref{eq:deltaav}) and (\ref{eq:deltacov})
if $A$ and $Q$ are interpreted as random variables.
To obtain the exponential decay of the covariance we
require that the probability distribution of the random momentum $Q$
is a Lorentzian,
 \begin{equation}
 {\cal{P}}_{Q} = \frac{\xi }{\pi } 
 \frac{ 1}{  ( Q \xi )^{2} + 1 }
 \label{eq:PQdef}
 \; ,
 \end{equation}
or equivalently for the random energy shift
$\eta$ defined in Eq.\ (\ref{eq:etadef}),
 \begin{equation}
 {\cal{P}}_{\eta}  
 = \frac{\gamma}{ \pi } \frac{ 1 }{ \eta^2 + {\gamma}^2 }
 \; ,
 \label{eq:peps}
 \end{equation}
with 
 \begin{equation}
 \gamma = \frac{v_F}{2 \xi}
 \; .
 \label{eq:gammadef}
 \end{equation}
The random variable $A$ should be distributed such that  
 \begin{eqnarray}
 \langle  A \rangle_A & = & 0 
 \label{eq:Afirstmom}
 \; ,
 \\
 \langle | A |^2 \rangle_A & = & \Delta_s^2
 \label{eq:Asecondmom}
 \; ,
 \end{eqnarray}
where $\langle \ldots \rangle_A$ denotes averaging over the
probability distribution of $A$.
From Eqs.\ (\ref{eq:PQdef}) to (\ref{eq:Asecondmom}) it is then easy to show
that the first two moments of
the distribution of $\Delta ( x )$ are
indeed given by Eqs.\ (\ref{eq:deltaav}) and (\ref{eq:deltacov}).
Note that Eqs.\ (\ref{eq:Afirstmom}) and (\ref{eq:Asecondmom})
include the cases of pure phase and pure amplitude
fluctuations.
To describe pure phase fluctuations we choose $A = \Delta_s e^{i \varphi}$,
where
the phase $\varphi$ is uniformly distributed in the interval
$[0, 2 \pi )$.  
\vspace{3.6cm}
\noindent
Then
 \begin{equation}
 \langle  \ldots \rangle_{A}^{\rm ph} =  \int_{0}^{2 \pi}
 \frac{ d \varphi}{2 \pi }  \ldots
 \; .
 \label{eq:phasemeasure}
 \end{equation}
Since physical quantities should be independent of the constant phase
$\varphi$ and therefore should only depend on $|A|$, the process of
averaging amounts to replacing $|A|$ by $\Delta_s$. 
To take into account amplitude fluctuations we follow
Sadovskii \cite{Sadovskii74,Sadovskii79} and choose
a Gaussian distribution for the real and imaginary parts of $A$,
\begin{equation}
 \langle  \ldots \rangle_A^{\rm am} = 
\int_{-\infty}^{\infty} \frac{ d {\rm Re}A \; d {\rm Im} A}{ \pi \Delta_s^2}
 e^{ - | A |^2 / \Delta_s^2 } \ldots
 \; .
 \label{eq:ampmeasure}
 \end{equation}
The disorder averaging of any functional
${\cal{F}} \{ \Delta ( x ) \}$ is defined by
 \begin{equation}
 \langle {\cal{F}} \{ \Delta (x ) \}
 \rangle  \equiv 
 \left\langle
 \int_{- \infty}^{\infty} d Q \,
 {\cal{P}}_Q \,
 {\cal{F}} \{ A e^{i Q x } \} \right\rangle_{A}
 \; .
 \label{eq:avdef2}
 \end{equation}
What is the physical meaning of an order parameter
of the form (\ref{eq:deltasimple})?
In a superconductor  
such  an order parameter describes a state
with a uniform superflow \cite{deGennes66}.
The gauge transformation (\ref{eq:gaugetrafo})
corresponds to choosing a coordinate system where
the superflow vanishes; $\eta$
is the associated energy shift. 
A more detailed physical justification for 
such a spatially constant random energy shift
$\eta$ in the normal state of the cuprate superconductors
has been given by Franz and Millis \cite{Franz98}: they
pointed out that within a semi-classical approximation the effect
of the quasi-static fluctuations of the phase of the 
order parameter field $\Delta (x )$ can be described by
such an energy shift $\eta$. 
Franz and Millis \cite{Franz98} also presented a perturbative calculation of the
probability distribution ${\cal{P}}_{\eta}$ of $\eta$, using earlier results by
Emery and Kivelson \cite{Emery95}. 
Because  in Ref. \cite{Franz98}  a cumulant expansion
of ${\cal{P}}_{\eta}$  was truncated at the second order,
the form of ${\cal{P}}_{\eta}$ was found to be Gaussian by
construction.
However, there are certainly non-Gaussian corrections to the
form of ${\cal{P}}_{\eta}$ given in Ref. \cite{Franz98}.
Our assumption that the distribution of $\eta$ is
a Lorentzian of width $\gamma$ is therefore not in contradiction
to the work of Ref. \cite{Franz98}. Obviously,
our parameter $\gamma$ in Eq.\ (\ref{eq:gammadef}) is the analog
of the parameter $W$ introduced in Eq.\ (9) of Ref. \cite{Franz98}.
Note, however, that Franz and Millis \cite{Franz98} did not consider
amplitude
fluctuations of the order parameter, which are described by our second
random variable $A$. As noted above, Gaussian amplitude fluctuations 
with a probability distribution given by Eq.\ (\ref{eq:ampmeasure})
have been
studied many years ago by Sadovskii \cite{Sadovskii74}. Thus, in the present
work we combine the models introduced by Sadovskii \cite{Sadovskii74}
and by Franz and Millis \cite{Franz98} such that we take both amplitude
and phase fluctuations into account and still obtain an exactly
solvable model.

In the following section we shall calculate a number of 
physical quantities for this model exactly and confirm
the intuitive picture \cite{Emery95,Franz98} that
phase fluctuations fill in the gap at the Fermi energy and render the
system metallic.

\section{Calculation of physical quantities}

\subsection{Single-particle Green's function and spectral function}

Because  $\langle A \rangle = 0$, it follows from
Eq.\ (\ref{eq:Gqqres}) that the off-diagonal elements of the
disorder averaged Green's function vanish, and that
the diagonal elements are
 \begin{equation}
   \langle
 {\cal{G}}_{\alpha \alpha} ( q , q^{\prime} , \omega  )
 \rangle
 = 2 \pi \delta ( q - q^{\prime})
 G_{\alpha} ( q , \omega )
 \; ,
 \label{eq:Galphadef}
 \end{equation}
where 
 \begin{equation}
 G_{\alpha} ( q , \omega ) =
 \left\langle 
 \frac{ \omega - 2 \eta + \alpha v_F q }{
 [ \omega - 2 \eta + \alpha v_F q ][ \omega - \alpha v_F q ]
 - | A |^2 }
 \right\rangle
 \; .
 \label{eq:Galphaav}
 \end{equation}
Here, $\alpha = +$ refers to ${\cal{G}}_{11}$, and
$\alpha =-$ refers to ${\cal{G}}_{22}$. 
The averaging over the Lorentzian distribution (\ref{eq:peps}) 
of the random energy shift
$\eta $ can be performed analytically, 
 \begin{equation}
 G_{\alpha} ( q , \omega + i 0^{+}) =
 \Biggl\langle 
 \frac{1}{ \omega - \alpha v_F q  -
 { \displaystyle 
 \frac{ | A |^2}{ \omega + \alpha v_F q
 +   i \frac{v_F}{\xi} }}
 } \Biggr\rangle_{A}
 \; ,
 \label{eq:GAA}
 \end{equation}
where
$\langle \ldots \rangle_A$ denotes averaging over the probability
distribution of $A$. In the case of pure phase fluctuations, as described
by Eq.\ (\ref{eq:phasemeasure}), this averaging is trivial, so that
 \begin{equation}
 G_{\alpha}^{\rm ph} ( q , \omega + i 0^{+}) =
 \frac{1}{ \omega - \alpha v_F q - 
 \Sigma_{\alpha}^{\rm ph} ( q , \omega + i 0^{+})
 }
 \label{eq:Gsigmares}
 \; ,
 \end{equation}
with the self-energy given by
 \begin{equation}
 \Sigma_{\alpha}^{\rm ph} ( q , \omega + i 0^{+}) =
 \frac{ \Delta_s^2 }{ \omega + \alpha v_F q
 +   i \frac{v_F}{\xi} }
 \; .
 \label{eq:sigmaphase}
 \end{equation}
Eq.\ (\ref{eq:sigmaphase}) agrees precisely with the lowest order
Born approximation, which was used in the seminal work by Lee, Rice,
and Anderson \cite{Lee73}. We have thus found a special probability
distribution of $\Delta ( x)$ where the lowest order Born
approximation for the average single-particle Green's function is
exact: the order parameter is in this case of the form
$\Delta ( x) = \Delta_s e^{ i Q x + i \varphi}$, where
$Q$ has a Lorentzian distribution of width $1/\xi$, and
the random phase $\varphi$ merely assures $\langle \Delta(x) \rangle
=0$, but due to gauge invariance does not affect any physical quantities.
 
On the other hand, if in addition to phase fluctuations
also  amplitude fluctuations are important,  
there are corrections to the Born approximation.
For Gaussian amplitude fluctuations given by
Eq.\ (\ref{eq:ampmeasure})
we find  after substituting
$t = | A |^2 / \Delta_s^2$  
 \begin{eqnarray}
 G_{\alpha}^{\rm ph+am} ( q , \omega + i 0^{+}) & = &
 \nonumber
 \\
 & & \hspace{-35mm} 
 \int_0^{\infty} dt  \frac{ e^{-t}}{ \omega - \alpha v_F  q 
 - 
{\displaystyle
 \frac{ t  \Delta_s^2}{ \omega + \alpha v_F q 
 +   i \frac{v_F}{  \xi} } } }
 \; .
 \label{eq:Gaaint}
 \end{eqnarray}
Recently Kuchinskii and Sadovskii \cite{Kuchinskii99}
arrived precisely at Eq.\ (\ref{eq:Gaaint}) within a
diagrammatic attempt to
estimate the accuracy of the method developed in Ref. \cite{Sadovskii79}
for Gaussian disorder. 
For a better comparision with Sadovskii's Green's function calculated
in Ref.\ \cite{Sadovskii79}, 
let us  represent Eq.\ (\ref{eq:Gaaint}) as a continued fraction.

\newpage



\noindent
Expressing the integral on the right-hand side of Eq.\ (\ref{eq:Gaaint})
in terms of the  incomplete
$\Gamma$-function and using the known continued fraction
expansion of this function \cite{Gradshteyn80}, we obtain for the
self-energy
 {
   \begin{eqnarray}
 \Sigma_{\alpha}^{\rm ph+am} ( q , \omega + i 0^{+})
  & = & 
 \nonumber
 \\
 & & \nonumber
 \\
 & & \hspace{-22mm} 
  \frac{\Delta_s^2}{
 \displaystyle 
  \omega + \alpha v_F q +  i \frac{v_F}{\xi}
 -  \frac{\Delta_s^2}{ \displaystyle   
  \omega - \alpha v_F q 
  -  \frac{2 \Delta_s^2}{ \displaystyle
 \omega + \alpha v_F q +  i \frac{v_F}{\xi} 
 -  \frac{ 2 \Delta_s^2}{\displaystyle
  \omega - \alpha v_F q
 - \frac{3 \Delta_s^2}{\displaystyle
  \omega + \alpha v_F q +  i \frac{v_F}{\xi}
 - \ldots }}}}} 
 \; .
 \nonumber
 \\ 
 & &
 \label{eq:sigmares}
 \end{eqnarray}
 }
For the same model with Gaussian disorder
the algorithm due to Sadovskii \cite{Sadovskii79} 
produces the continued fraction expansion
 {
 \begin{eqnarray}
 \Sigma_{\alpha}^{ \rm Sadovskii} ( q , \omega + i 0^{+})
  & = &
 \nonumber
 \\
 & & \nonumber
 \\
 & & \hspace{-32mm}
  \frac{\Delta_s^2}{ \displaystyle
  \omega + \alpha v_F q +  i \frac{v_F}{\xi}
 -  \frac{\Delta_s^2}{ \displaystyle
  \omega - \alpha v_F q + 2 i \frac{v_F}{\xi}
  -   \frac{2 \Delta_s^2}{ \displaystyle
 \omega + \alpha v_F q + 3 i \frac{v_F}{\xi}
 -  \frac{2 \Delta_s^2}{ \displaystyle
  \omega - \alpha v_F q + 4 i \frac{v_F}{\xi}
 - \frac{3 \Delta_s^2}{\displaystyle
  \omega + \alpha v_F q + 5 i \frac{v_F}{\xi}
 - \ldots}}}}}
 \nonumber
 \; .
 \\
 & &
 \label{eq:sigmasadres}
 \end{eqnarray}
 }
Note that only the first two lines in 
Eqs.\ (\ref{eq:sigmares}) and (\ref{eq:sigmasadres}) agree.
Kuchinskii and Sadovskii argue in Ref. \cite{Kuchinskii99} that
the true behavior of the Green's function for Gaussian disorder lies
somewhat in between Eqs.\ (\ref{eq:sigmares}) and (\ref{eq:sigmasadres}).
%
%
%
%
In our model,
the coexistence of amplitude fluctuations 
with phase fluctuations (which are related to our random energy shift $\eta$)
generates a completely new feature 
in the average spectral function.
The latter is related to the av-
\newpage

\mbox{}

\begin{figure}
\begin{center}
\psfrag{A}{$\Delta_s \langle \rho ( k_F , \omega ) \rangle $}
\psfrag{om}{$\bar{\omega}$}
\psfrag{0}{0}
\psfrag{0.5}{0.5}
\psfrag{1}{1}
\psfrag{2}{2}
\psfrag{-2}{-2}
\psfrag{-1}{-1}
\epsfig{file=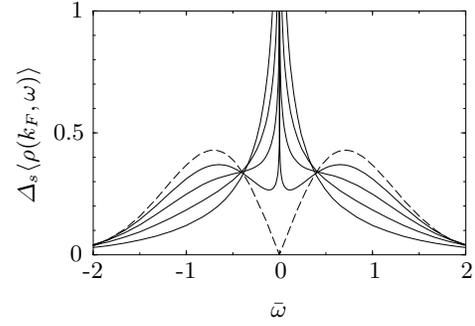,width=6cm}
\end{center}
\caption{Average spectral function  
$ \langle \rho ( k_F , \omega ) \rangle^{\rm ph+am}$
as a function of $\bar{\omega} = \omega / \Delta_s$ 
for $\bar{\gamma} = 0$ (dashed line, see Eq.\ (\ref{eq:specinfty})),
and $\bar{\gamma} = 0.1, 0.25, 0.5, 1 $ (see Eq.\ (\ref{eq:avspecres})).
For any finite $\bar{\gamma}$
there is a  logarithmic singularity at $\omega =0$, 
which acquires more weight as $\bar{\gamma}$ increases. 
} 
\label{fig:specq0}
\end{figure}

\begin{figure}
\begin{center}
\psfrag{A}{$\Delta_s \langle \rho ( k_F + q , \omega ) \rangle $}
\psfrag{om}{$\bar{\omega}$}
\psfrag{0}{0}
\psfrag{0.5}{0.5}
\psfrag{1}{1}
\psfrag{2}{2}
\psfrag{-2}{-2}
\psfrag{-1}{-1}
\epsfig{file=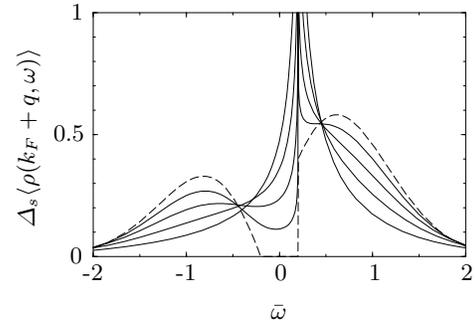,width=6cm}
\end{center}
\caption{Average spectral function  
$ \langle \rho ( k_F + q , \omega ) \rangle^{\rm ph+am}$
as a function of $\bar{\omega} = \omega / \Delta_s$
for $v_F q / \Delta_s = 0.2$.
The dashed line corresponds to $\bar{\gamma}=0$ 
(see Eq.\ (\ref{eq:specinfty})), while
the other curves correspond to $\bar{\gamma} = 0.1, 0.25, 0.5,1 $ (see 
Eq.\ (\ref{eq:avspecres})).
} 
\label{fig:specq02}
\end{figure}

\newpage
\noindent
erage Green's function via
\begin{equation}
 2 \pi \delta ( q - q^{\prime} ) \langle \rho ( \alpha k_F + q  , \omega )  \rangle
 =
 - \frac{1}{\pi} \,
 {\rm Im} \,   \langle
 {\cal{G}}_{\alpha \alpha} ( q , q^{\prime} , \omega + i 0^{+} )
 \rangle
 \; ,
 \label{eq:Adef}
 \end{equation}
Using Eq.\ (\ref{eq:Gaaint}) we find
 \begin{eqnarray}
 \langle \rho ( \alpha k_F + q , \omega ) \rangle^{\rm ph+am} & = & 
\nonumber 
\\
 &  & \hspace{-40mm}
 \frac{2 \bar{\gamma}}{\pi \Delta_s}
 \int_0^{\infty} d t \frac{ t e^{-t}}{ ( t - \bar{\omega}^2 +
   \bar{q}^2 )^2 + 4 \bar{\gamma}^2 ( \bar{\omega} - 
 \alpha \bar{q})^2 }
 \; ,
 \label{eq:avspecres}
 \end{eqnarray}
where $\bar{q} = v_F q / \Delta_s$, $\bar{\omega} = \omega / \Delta_s$,
and $\bar{\gamma} = v_F / (2 \Delta_s \xi)$.
Representative results for different values
of $\bar{\gamma}$ are shown in
Figs. \ref{fig:specq0} and \ref{fig:specq02}.
The dashed line is the spectral function 
for $\bar{\gamma}=0$ (i.e. without phase fluctuations),
which is easily calculated analytically,
 \begin{eqnarray}
  \langle \rho ( \alpha k_F + q , \omega ) \rangle^{\rm am}
 & = & 
 \nonumber
 \\
 &  & \hspace{-20mm} 
\Delta_s^{-1} \Theta ( {\bar{\omega}}^2 - {\bar{q}}^2 )
 |  \bar{\omega}  + \alpha \bar{q} |
 e^{ - ( \bar{\omega}^2 - \bar{q}^2 )}
 \; .
 \label{eq:specinfty}
 \end{eqnarray}
The important point is now that 
for any finite $\bar{\gamma}$ the spectral function exhibits 
a logarithmic singularity at $\omega = \alpha v_F q$. 
In the vicinity of this singularity the leading behavior of the spectral function
can be calculated analytically. In the regime
 \begin{equation}
 | \omega - \alpha v_F q | \ll {\rm min} 
 \left\{   \frac{ \Delta_s^2 \xi}{v_F} , 
 \frac{ \Delta_s^2 }{|\omega + \alpha v_F q | }
 \right\}
 \label{eq:regime}
 \end{equation}
the integral in Eq.\ (\ref{eq:avspecres})
can be approximated by
\begin{eqnarray}
 \langle \rho ( \alpha k_F + q  , \omega )  \rangle^{\rm ph+am}
 & \sim & \frac{ 2 \bar{\gamma}}{\pi \Delta_s}
 \ln \left[ \frac{ 1}{ 2 \bar{\gamma}  | \bar{\omega} - \alpha \bar{ q} |}
 \right]
 \nonumber
 \\
 &  &  \hspace{-20mm} =
 \frac{ v_F }{\pi \Delta_s^2 \xi }
 \ln \left[ \frac{ \Delta_s^2 \xi }{ v_F | \omega - \alpha v_F q |}
 \right]
 \; .
 \label{eq:speclogdimless}
 \end{eqnarray}
Thus, the interplay between phase fluctuations (described by our 
random phase factor $e^{i Q x}$) and amplitude fluctuations
(described by random fluctuations of $ | A | $)
gives rise to a logarithmic  singularity at the bare
energy of the electron. Note that such a singularity
is  weaker than the algebraic singularities that are typically
found in the spectral function of a Luttinger liquid.
Of course, such a weak singularity cannot be called a quasi-particle
peak. 
It is important to point out that
in the presence of amplitude fluctuations alone or 
phase fluctuations alone 
such a logarithmic singularity does not exist. 
Recall that for pure phase fluctuations
our model has the same spectral function 
as predicted by the Born approximation for
the self-energy \cite{Lee73},
while for pure amplitude fluctuations our model
reduces to the model discussed by Sadovskii in Ref. \cite{Sadovskii74}.
Note also that the approximate spectral function 
produced by Sadovskii's algorithm \cite{Sadovskii91,McKenzie96}
for Gaussian disorder
with a finite correlation length
does not exhibit any logarithmic singularities.
Whether an exact calculation of the spectral function
for more realistic probability distributions could confirm this result
or not remains an open question. 
 
From Fig.\ \ref{fig:specq02} it is clear that the line-shape
of the spectral function in the vicinity of the singularity is rather
broad and asymmetric. Such a behavior has recently been seen 
in the photoemission spectra of a one-dimensional band-insulator \cite{Vescoli00}.


\subsection{Average density of states}
\begin{figure}
\begin{center}
\psfrag{om}{$\bar{\omega}$}
\psfrag{DOS}{$\hspace{-2mm}\langle \rho ( \omega ) \rangle / \rho_0$}
\psfrag{0}{0}
\psfrag{0.5}{0.5}
\psfrag{1.5}{1.5}
\psfrag{0}{0}
\psfrag{1}{1}
\psfrag{2}{2}
\psfrag{3}{3}
\psfrag{4}{4}
\epsfig{file=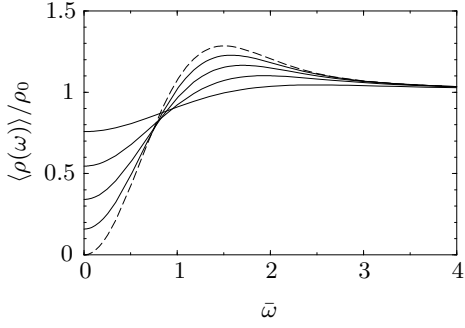,width=6cm}
\end{center}
\caption{Average DOS 
$\langle \rho ( \omega ) \rangle^{\rm ph+am}$
[see Eq.\ (\ref{eq:rhoampres})] as a function of
$\bar{\omega} = \omega / \Delta_s$ for 
$\bar{\gamma} =0$ (dashed line), and $\bar{\gamma} = 0.1, 0.25, 0.5,
1$. 
For smaller $\bar{\gamma}$ the pseudogap becomes deeper.}
\label{fig:rhoomega}
\end{figure}

\begin{figure}
\begin{center}
\psfrag{g}{$\bar{\gamma}$}
\psfrag{R}{$\hspace{-2mm}\langle \rho ( 0 ) \rangle / \rho_0$}
\psfrag{0}{0}
\psfrag{0.5}{0.5}
\psfrag{0}{0}
\psfrag{1}{1}
\psfrag{2}{2}
\psfrag{3}{3}
\psfrag{4}{4}
\epsfig{file=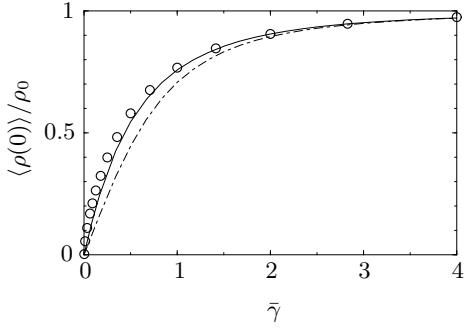,width=6cm}
\end{center}
\caption{Solid line: numerical evaluation of
$R (\bar{\gamma}) = \langle \rho ( 0) \rangle^{\rm ph+am} / \rho_0$ 
as a function of
$\bar{\gamma} = v_F / ( 2 \Delta_s \xi )$ (see Eq.\ (\ref{eq:Rgdef})).
Dashed-dotted line: the same quantity without amplitude 
fluctuations  (see Eq.\ (\ref{eq:rhophaseres})),
which amounts to calculating the average DOS from the
self-energy in first order Born approximation, as was done by 
Lee, Rice, and Anderson \cite{Lee73}.
The circles are  numerical  results for Gaussian
disorder, obtained via the exact numerical algorithm of Ref.\
\cite{Bartosch99b}. 
}
\label{fig:rhozero}
\end{figure}

The average DOS is defined by
 \begin{equation}
 \langle \rho ( \omega ) \rangle = 
 - \frac{1}{\pi} \,
 {\rm Im}\,  {\rm Tr}\,  \langle
 {\cal{G}} ( x , x , \omega + i 0^{+} )
 \rangle
 \; .
 \label{eq:localdos}
 \end{equation}
Performing the $q$-integration in Eq.\ (\ref{eq:GAA}) we find
 \begin{equation}
 {\rm Tr}\,  \langle
 {\cal{G}} ( x , x , \omega + i 0^{+} )
 \rangle
  = - \frac{1}{v_F} \left\langle
 \frac{ \omega + i \gamma}{ \sqrt{ | A |^2 - ( \omega + i \gamma )^2 }
   } \right\rangle_A  
 \; ,
 \label{eq:GAAxx}
 \end{equation}
where  $ \gamma $ is given in Eq.\ (\ref{eq:gammadef}) and
$\sqrt{z}$ denotes the principal branch of the square root, with the
cut at the negative real axis.
Note that phase fluctuations simply generate an imaginary
shift $i \gamma$ to the frequency in Eq.\ (\ref{eq:GAAxx}). 
In the absence of amplitude fluctuations 
(see Eq.\ (\ref{eq:phasemeasure}))
we may replace
$|A| \rightarrow \Delta_s$ in Eq.\ (\ref{eq:GAAxx}), so that
we obtain for the average DOS
 \begin{equation}
 \frac{ \langle \rho ( \omega ) \rangle^{\rm ph}}{\rho_0} =
 {\rm Im }\, \frac{ z }{ \sqrt{ 1 - z^2}}
 \; ,
 \label{eq:rhophaseres}
 \end{equation}
where we have defined
 \begin{equation}
 z = \frac{ \omega + i \gamma}{\Delta_s} =
 \bar{\omega} + i \bar{\gamma}
 \label{eq:zdef}
 \; .
 \end{equation}
Eq.\ (\ref{eq:rhophaseres}) agrees exactly with the perturbative result
by  Lee, Rice, and Anderson \cite{Lee73}.
For $\omega = 0$ we recover Eq.\ (\ref{eq:rhoBorn}).
On the other hand, in the presence of additional 
Gaussian amplitude fluctuations,
with probability distribution given by Eq.\ (\ref{eq:ampmeasure}), we obtain
 \begin{equation}
 \frac{\langle \rho ( \omega ) \rangle^{\rm ph+am}}{\rho_0} =
 {\rm Im }  \int_0^{\infty} d t  \frac{ e^{-t} z }{ \sqrt{ t - z^2}}
 \; .
 \label{eq:rhoampres}
 \end{equation}
A numerical evaluation of 
Eq.\ (\ref{eq:rhoampres}) is shown in Fig.\ \ref{fig:rhoomega}.
For $\gamma = 0$ the integral in Eq.\ (\ref{eq:rhoampres}) 
can be done analytically and reduces to the result
obtained by Sadovskii \cite{Sadovskii74}, which 
does not contain phase fluctuations.
In this case the DOS vanishes quadratically for small frequencies,
 \begin{equation} 
 \frac{\langle \rho ( \omega ) \rangle^{\rm am}}{\rho_0}  \sim
 2  {\bar\omega  }^2 
 \; \; , \; \; | \bar\omega | \ll 1
 \; .
 \end{equation}
For any finite $\xi$ the DOS at the Fermi energy (i.e. at $\omega =
0$) is finite. From Eq.\ (\ref{eq:rhoampres}) we find
 \begin{equation}
 \frac{\langle \rho ( 0 ) \rangle^{\rm ph+am} }{\rho_0}= 
 R ( \bar{\gamma} ) 
 \; ,
 \label{eq:rhozero}
 \end{equation}
with
 \begin{equation}
 R ( \bar{\gamma} ) = \bar{\gamma} \int_0^{\infty} dt  
 \frac{ e^{-t}}{ \sqrt{t + \bar{\gamma}^2}}
 \; .
 \label{eq:Rgdef}
 \end{equation}
A numerical evaluation of $R ( \bar\gamma )$ is shown
in Fig.\ \ref{fig:rhozero}.
For small and large $\bar\gamma$ 
we obtain to leading order
 \begin{equation}
 R ( \bar{\gamma} ) \sim 
 \left\{
 \begin{array}{ll}
 \sqrt{\pi} \bar{\gamma} & \; , \; \bar\gamma \ll 1 \\
 1 & \; , \; \bar{\gamma} \gg 1
 \end{array}
 \right.
 \; .
 \label{eq:Rasym}
 \end{equation}
For large $\xi$  the DOS at the Fermi energy is
 \begin{equation}
 \langle \rho ( 0 ) \rangle^{\rm ph+am}  \sim
 \frac{\sqrt{\pi}}{2 \pi  \Delta_s \xi}
 \;, \; \; v_F \xi \gg \Delta_s
 \; ,
 \label{eq:rhozerosmall}
 \end{equation}
which should be compared  with the
result obtained within the Born approximation, see
Eq.\ (\ref{eq:rhoBornsmall}),
 \begin{equation}
 \langle \rho ( 0 ) \rangle^{\rm pert} =
 \langle \rho ( 0 ) \rangle^{\rm ph} \sim
 \frac{1}{2 \pi \Delta_s \xi}
 \; .
 \label{eq:rhoBornsmall2}
 \end{equation}
Hence, Gaussian amplitude fluctuations  increase the value of the
DOS at the Fermi energy
as compared with pure phase fluctuations.
However, from Fig.\ \ref{fig:rhozero} it is
evident that the qualitative behavior of the
DOS is correctly predicted by a model with pure phase fluctuations,
which exactly reproduces the perturbative result \cite{Lee73}.
Let us emphasize that this is not the case if $\Delta ( x)$
has a Gaussian distribution: the prediction of lowest order
perturbation theory, $\langle \rho (0) \rangle \propto \xi^{-1}$,
is in disagreement with the exact numerical result
for Gaussian disorder, $\langle \rho (0) \rangle \propto \xi^{-0.64}$
(see Eq.\ (\ref{eq:rhoGauss})). 
We thus conclude that
the behavior of the average 
DOS at the Fermi energy of the FGM in one dimension is non-universal
and sensitive to the detailed form
of the probability distribution of $\Delta (x)$.

\subsection{Lyapunov exponent and localization length}
\label{subsec:localization}
\begin{figure}
\begin{center}
\psfrag{om}{$\bar{\omega}$}
\psfrag{IL}{$\hspace{-6mm} \ell^{-1}(\omega ) / v_F^{-1} \Delta_s$}
\psfrag{0}{0}
\psfrag{0.5}{0.5}
\psfrag{0}{0}
\psfrag{1}{1}
\psfrag{2}{2}
\psfrag{3}{3}
\psfrag{4}{4}
\epsfig{file=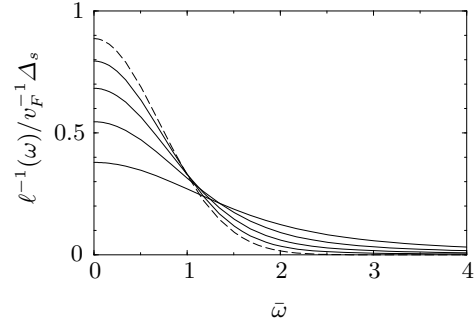,width=6cm}
\end{center}
\caption{Inverse localization length  
$\ell^{-1}(\omega )^{\rm ph+am} / v_F^{-1} \Delta_s$
[see Eq.\ (\ref{eq:Lyapunovresamp})] as a function of
$\bar{\omega} = \omega / \Delta_s$ for 
$\bar{\gamma} =0$ (dashed line), and $\bar{\gamma} = 0.1, 0.25, 0.5,
1$. 
}
\label{fig:invloc-omega}
\end{figure}

\begin{figure}
\begin{center}
\psfrag{g}{$\bar{\gamma}$}
\psfrag{P}{$\hspace{-2mm}v_F /  \Delta_s \ell (0)$}
\psfrag{0.5}{0.5}
\psfrag{0}{0}
\psfrag{1}{1}
\psfrag{2}{2}
\psfrag{3}{3}
\psfrag{4}{4}
\epsfig{file=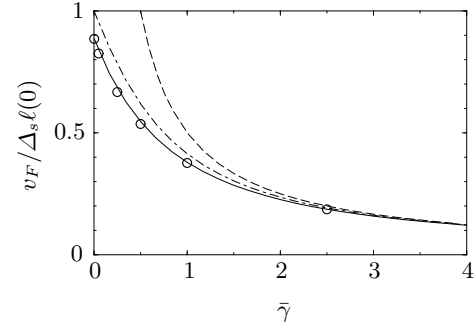,width=6cm}
\end{center}
\caption{
Inverse localization length $P  (\bar{\gamma}) = v_F 
/  \Delta_s \ell (0)$ at the Fermi energy for different
types of disorder.
Solid line: phase and amplitude fluctuations, 
see Eqs.\ (\ref{eq:loczerores}) and (\ref{eq:Pinfzero}); 
dashed-dotted line: phase fluctuations, 
see Eq.\ (\ref{eq:Lyapunovresphase});  
dashed line: extrapolation of white noise limit
$P ( \bar{\gamma}) = 1 / ( 2 \bar{\gamma})$, see
Eq.\ (\ref{eq:whitenoiseloc}).
The circles are numerical  results for Gaussian
disorder, obtained via an exact numerical algorithm
\cite{Bartosch00}.
}
\label{fig:P}
\end{figure}

Since the energy dispersion of the FGM is linear, the Schr\"{o}dinger equation
$\hat{H}_x \psi_{\omega } ( x ) = \omega \psi_{\omega } ( x )$ is a system of linear first order
differential equations. Fixing 
the two-component wave-function 
$\psi_{\omega}  ( x )$ arbitrarily 
at one space point $x_0$ therefore constitutes
the wave-function at all points $x$.
In a disordered system, the Lyapunov exponent $\kappa ( \omega )$
characterizes the  
exponential growth of the magnitude of the wave-function 
at large distances $| x-x_0|$ \cite{Lifshits88},
 \begin{equation}
 | \psi_{\omega } ( x ) | 
 \sim  | \psi_{\omega} ( x_0 ) |  \exp [ \kappa ( \omega ) | x - x_0 | ]
 \; .
 \label{eq:lyapunovdef}
 \end{equation}
Strictly speaking, the Lyapunov exponent is defined by the limit
$|x-x_0| \to \infty$ of this equation and assumes a
certain value with probability one \cite{Lifshits88}.
In one dimension the inverse of the
Lyapunov exponent can be identified with the {\it mean} localization
length.
According to the Thouless formula the {\it{mean}} localization length
$\ell ( \omega )$ can be obtained from the real part
of the disorder-averaged single-particle Green's function.
Originally the Thouless formula was derived for a one-band model with quadratic energy 
dispersion \cite{Thouless72}, but it can be shown to hold also
for the  FGM, where it can be written as \cite{Hayn87,Bartosch00}
 \begin{equation}
 \frac{\partial}{\partial \omega} \frac{1}{\ell ( \omega )}
 =   {\rm Re} {\rm Tr}
 \langle 
 {\cal{G}} ( x , x , \omega + i 0^{+} ) \rangle
 \; .
 \label{eq:Thouless}
 \end{equation}
Integrating the Thouless formula for Eq.\ (\ref{eq:GAAxx}), we obtain 
 \begin{equation}
 \frac{v_F}{\ell ( \omega )}  = {\rm Re} \left\langle 
  \sqrt{ | A |^2 - ( \omega + i \gamma )^2 } \right\rangle_A  - \gamma
 \; ,
 \label{eq:Lyapunovres}
 \end{equation}
where the constant of integration is uniquely determined by the
requirement $\lim_{\omega \rightarrow \infty } \ell^{-1} ( \omega ) = 0$.
For pure phase fluctuations Eq.\ (\ref{eq:Lyapunovres}) reduces to
  \begin{equation}
 \frac{v_F}{\Delta_s \ell ( \omega )^{\rm ph}} = {\rm Re}\,  
  \sqrt{ 1  - (\bar \omega + i\bar \gamma)^2 } - \bar{\gamma} 
 \; ,
 \label{eq:Lyapunovresphase}
 \end{equation}
while with additional Gaussian amplitude fluctuations
  \begin{equation}
  \frac{v_F}{\Delta_s \ell ( \omega )^{\rm ph+am}}  =
 {\rm Re} \left[
 \int_0^{\infty} dt e^{-t}  
  \sqrt{ t  - (\bar \omega + i\bar \gamma)^2 } \right]- \bar{\gamma}
 \; .
 \label{eq:Lyapunovresamp}
 \end{equation}
A plot of the inverse localization length $\ell^{-1}(\omega)^{\rm
  ph+am}$ is given in Fig.\ \ref{fig:invloc-omega}.
For $\gamma \rightarrow 0$ only amplitude fluctuations
are left, and Eq.\ (\ref{eq:Lyapunovresamp}) reduces to
 \begin{equation}
 \frac{v_F}{\Delta_s \ell ( \omega )^{\rm am} }
 = \frac{ \sqrt{\pi}}{2} e^{- \bar{\omega}^2}
 \; \; , \; \; \bar{\gamma} \rightarrow 0
 \; .
 \label{eq:Thres2}
 \end{equation} 
In the presence of phase and amplitude fluctuations
the general expression (\ref{eq:Lyapunovresamp})
simplifies at the Fermi energy to
 \begin{equation}
 \frac{v_F}{\Delta_s \ell ( 0 )^{\rm ph+am} }
 \equiv  P ( \bar{\gamma} )
 \;,
 \label{eq:loczerores}
 \end{equation}
where the dimensionless function $P ( \bar{\gamma})$ is given by
  \begin{equation}
 P ( \bar{\gamma} ) =  
 \int_{0}^{\infty}  dt e^{-t}
 \left[  \sqrt{ t + \bar{\gamma}^2 } - \bar{\gamma} \right]
 \label{eq:Pinfzero}
 \; .
 \end{equation}
A comparison  of Eq.\ (\ref{eq:Pinfzero})
with the corresponding expression obtained from Eq.\
(\ref{eq:Lyapunovresphase}) 
for phase fluctuations
is shown in Fig.\ \ref{fig:P}.
For small  and large
$\bar{\gamma}$ the leading behavior is
 \begin{equation}
 P ( \bar{\gamma} ) \sim
 \left\{
 \begin{array}{ll}
 {\sqrt{\pi}}/{2}
 & \; , \; \bar{\gamma} \ll 1  \\
 { 1}/({2 {\bar{\gamma}}})  & \; , \; \bar{\gamma} \gg 1
 \end{array}
 \right.
 \label{eq:Pinftyasym}
 \; .
 \end{equation}
In the white noise limit $\xi \rightarrow 0$, $\Delta_s \rightarrow
\infty$ with $\Delta_s^2 \xi = {\rm{const}}$ only the behavior
of $P ( \bar{\gamma})$ for large $\bar{\gamma}$ matters, and
in this limit both Eq.\ (\ref{eq:Lyapunovresphase}) and
Eq.\ (\ref{eq:loczerores}) reduce to the known white-noise result
 \begin{equation}
 \frac{v_F}{ \ell ( 0 ) } = \frac{\Delta_s}{2 \bar{\gamma}} =
 \frac{\Delta_s^2 \xi}{ v_F}
 \;  , \;  \xi \rightarrow 0 \; {\rm with} \; \Delta_s^2 \xi = {\rm const}
 \; .
 \label{eq:whitenoiseloc}
 \end{equation}
An extrapolation of this white-noise result 
towards finite correlation lengths is shown as the dotted line
in Fig.\ \ref{fig:P}. 
Evidently, for large $\gamma$ the behavior of the localization length
becomes independent of the precise form of the probability distribution
of the disorder. 
For $\bar{\gamma } { \raisebox{-0.5ex}{$\; \stackrel{<}{\sim} \;$}} 1$
the localization length begins to deviate significantly from the
white-noise limit and approaches a finite value 
of the order of $ v_F / \Delta_s$
for $\bar{\gamma} \rightarrow 0$, the precise value of which
depends on the type of the disorder.
We emphasize that 
for a real order parameter the low-frequency
behavior of the localization length is dominated by
the Dyson singularity, so that in this case $1/ \ell ( 0) = 0$ 
for any finite value of $\bar{\gamma}$, see Refs. \cite{Hayn87,Bartosch00}.

To compare the localization length
of our exactly solvable toy model with phase and amplitude fluctuations
with the case where the distribution of $\Delta (x)$ is a Gaussian, 
we have evaluated the Thouless formula (\ref{eq:Thouless})
numerically for Gaussian colored noise with correlation length $\xi$,
using an algorithm \cite{Bartosch00} similar to the one developed in Ref.\
\cite{Bartosch99b}.
The numerical results for 
$v_F / ( \Delta_s \ell ( 0 ))$ are shown as the open circles
in Fig.\ \ref{fig:P}. In view of the simplicity
of our model the agreement with Eq.\ (\ref{eq:loczerores})
is quite spectacular. 
Hence, the localization length of our model with 
phase and amplitude fluctuations
is a very accurate approximation 
to  the localization length
of the FGM with Gaussian disorder.
The dashed line in Fig.\ \ref{fig:P} describes the localization length
for the case 
where we ignore amplitude fluctuations in our model, which is
equivalent to the perturbative result by Lee, Rice, and Anderson \cite{Lee73}.
The  agreement with the case of Gaussian disorder
is not so good, in particular in the pseudogap regime 
$\bar{\gamma } { \raisebox{-0.5ex}{$\; \stackrel{<}{\sim} \;$}} 1$.

\subsection{Average conductivity}
\label{subsec:conductivity}
\begin{figure}
\begin{center}
\psfrag{g}{$\bar{\gamma}$}
\psfrag{D}{$\hspace{-1mm} D, W_{\rm deloc}(0)$}
\psfrag{0}{0}
\psfrag{0.5}{0.5}
\psfrag{0}{0}
\psfrag{1}{1}
\psfrag{2}{2}
\psfrag{3}{3}
\psfrag{4}{4}
\epsfig{file=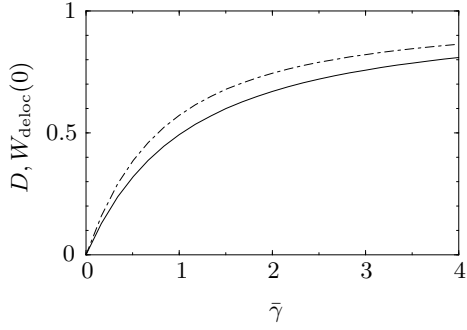,width=6cm}
\end{center}
\caption{
Solid line: dimensionless renormalization factor $D$ of the weight
of the Drude peak as a function of
$\bar{\gamma} = v_F / ( 2 \Delta_s \xi)$, see Eq.\ (\ref{eq:Ddef});
dashed-dotted line: probability $W_{\rm deloc} (0 )$ for finding delocalized states at the Fermi
energy, see Eq.\ (\ref{eq:Wres}).
}
\label{fig:DrudeP}
\end{figure}

The DOS and the spectral function 
[see Eqs.\ (\ref{eq:Adef}) and (\ref{eq:localdos})] 
involve only  the diagonal elements of the single-particle
Green's function. The simplest 
physical quantity which involves also the off-diagonal elements
of ${\cal{G}}$
is the average polarization 
$ \langle \Pi ( q , i \omega_m ) \rangle$,
which is given by
 \begin{eqnarray}
 2 \pi \delta ( q - q^{\prime} )
 \langle \Pi ( q , i \omega_m ) \rangle
 & = & - \frac{1 }{\beta} \sum_n \int \frac{ d p}{ 2 \pi} \int 
 \frac{ d p^{\prime}}{ 2 \pi} 
 \nonumber
 \\
 & & \hspace{-35mm} \times
 {\rm Tr} \langle
 {\cal{G}} ( p + q , p^{\prime}  + q^{\prime} , i \tilde{\omega}_{ n + m } )
 {\cal{G}} ( p^{\prime} , p , i \tilde{\omega}_{ n  } )
 \rangle
 \; .
 \label{eq:poldef}
 \end{eqnarray}
Here, $\beta$ is the inverse temperature, $\omega_m = 2 \pi m / \beta$ are
bosonic Matsubara frequencies and $\tilde{\omega}_n = 2 \pi ( n + \frac{1}{2} ) / \beta$
are fermionic ones.
Given the average polarization, the average
conductivity is easyly obtained from
 \begin{equation}
 \langle \sigma ( q , \omega ) \rangle =
  - e^2 \frac{  i  \omega }{q^2} 
 \langle \Pi ( q , \omega + i 0^{+} ) \rangle
 \; .
 \end{equation}
In this work we shall only consider
the real part of the conductivity at $q=0$,
 \begin{equation}
 {\rm Re}\, \langle \sigma ( \omega ) \rangle = \lim_{ q \rightarrow
   0} {\rm Re} \,
 \langle \sigma ( q , \omega ) \rangle
 = e^2  \omega \lim_{q \rightarrow 0} \frac{ 
 \langle {\rm Im}\, \Pi ( q , \omega 
 + i 0^{+} ) \rangle }{q^2}
 \; .
 \label{eq:sigmaprimedef}
 \end{equation}
Substituting Eq.\ (\ref{eq:Gqqres}) into Eq.\ (\ref{eq:poldef})  and performing the
Matsubara sum, we obtain for the average polarization
 \begin{eqnarray}
 \langle \Pi  ( q , i \omega_m ) \rangle & = &
 \nonumber 
 \\
 & & \hspace{-27mm}
 \left\langle
 - \int \frac{dp}{ 2 \pi}
 \frac{E_p E_{p+q} + \xi_p \xi_{p+q} + | A |^2}{2 E_p E_{p+q}}
 \nonumber
 \right.
 \\
 &  &  \hspace{-20mm} \times \left[ 
 \frac{ f ( E_p - \eta ) - f ( E_{p+q} - \eta ) }{ E_p - E_{p+q} - i \omega_m}
 \nonumber
 \right.
 \\
 & & \left. \hspace{-17mm} +
 \frac{ f ( E_p + \eta ) - f ( E_{p+q} + \eta ) }{ E_p - E_{p+q} + i \omega_m}
 \right]
 \nonumber
 \\
 &  & \hspace{-27mm} +
  \int \frac{dp}{ 2 \pi}
 \frac{E_p E_{p+q} - \xi_p \xi_{p+q} - | A |^2}{2 E_p E_{p+q}}
 \nonumber
 \\
 &  &  \hspace{-20mm} \times
 \left[
 \frac{ 1 - f ( E_p - \eta ) - f ( E_{p+q} + \eta ) }{ 
 E_p + E_{p+q} - i \omega_m}
 \nonumber
 \right. 
 \nonumber
 \\
 & & \left. \left. \hspace{-17mm} +
 \frac{ 1 - f ( E_p + \eta ) - f ( E_{p+q} - \eta ) }{ E_p + E_{p+q} + i \omega_m}
 \right]
 \right\rangle
 \label{eq:Pimatsubara}
 \; ,
 \end{eqnarray}
where we use the notation
$E_p = ( \xi_p^2 + | A |^2)^{1/2}$, $\xi_p = v_F p$
and $ f ( E ) = 1 / [e^{\beta E} + 1 ] $ is the Fermi-Dirac
function.
Setting $\eta = 0$ in Eq.\ (\ref{eq:Pimatsubara}) 
we recover  Eq.\ (2.10) of Ref. \cite{Sadovskii74}.
Expanding Eq.\ (\ref{eq:Pimatsubara}) for small $q$ and performing the
average over the Lorentzian distribution of $\eta$, we obtain
in the limit of zero temperature $(\beta \rightarrow \infty)$,
 \begin{eqnarray}
  {\rm Re}\, \langle \sigma ( \omega ) \rangle
  & =  &  \frac{   n e^2  }{m}
 \frac{\pi   }{\gamma}
 \left\langle
  \sqrt{  | A |^2 + \gamma^2  } - | A |  \right\rangle_A
 \delta ( \omega ) 
 \nonumber
 \\
 &  & \hspace{-15mm }+ \frac{  n e^2  }{m}
 \arctan \left( \frac{ | \omega |}{\gamma} \right)
 \left\langle \frac{ | A |^2 }{\omega^2} \frac{ \Theta ( \omega^2 - | A |^2 )}{
 \sqrt{ \omega^2 - | A |^2 }} \right\rangle_{A}
 \; ,
 \end{eqnarray}
where $ n/m \equiv v_F / \pi$ 
and ${\gamma}$ is defined in Eq.\ (\ref{eq:gammadef}).
For pure phase fluctuations the averaging over
the distribution of $A$ is trivial and
simply leads to the replacement $| A| \rightarrow \Delta_s$.
Then the conductivity exhibits a Drude peak with weight
given by $ \bar{\gamma}^{-1} ( \sqrt{ \Delta_s^2 + \bar{\gamma}^2} -
\Delta_s )$, which is
separated from a continuum at higher frequencies by a finite gap
$\Delta_s$. 
Gaussian amplitude fluctuations wash out the gap but do not
remove the Drude peak.
Averaging over the probability distribution
of the amplitude $ A $ given in Eq.\ (\ref{eq:ampmeasure}) 
we  obtain
 \begin{equation}
  {\rm Re}\, \langle \sigma ( \omega ) \rangle
  =  \frac{n e^2 }{m } 
  \left[
  \pi D ( \bar\gamma ) \delta  ( \omega )+ \frac{1}{\Delta_s}
  C ( \bar\gamma , \bar{\omega} )
  \right]
  \label{eq:ResigmaDC}
  \; ,
  \end{equation}
where we have used again the notation $\bar\gamma = \gamma / \Delta_s$,
$\bar{\omega } = \omega / \Delta_s$, and
the dimensionless functions $D ( \bar{\gamma} )$ and
$C ( \bar{\gamma} , \bar{\omega} )$ are
 \begin{equation}
  D  ( \bar{\gamma} )  
  =  \frac{1}{ \bar{\gamma} } 
  \int_0^{\infty} d t\, e^{-t} [ \sqrt{ t + \bar{\gamma}^2} - \sqrt{t} ]
 \; ,
 \label{eq:Ddef}
 \end{equation}
 \begin{equation}
 C ( \bar{\gamma} , \bar\omega ) =
 \arctan \left( \frac{ | \bar{\omega} | }{ \bar{\gamma} } \right)
 | \bar{\omega} | 
 \int_0^{1} dt e^{- \bar{\omega}^2 t} \frac{t}{\sqrt{ 1 - t }}
 \; .
 \label{eq:Cdef}
 \end{equation}
A graph of $D (\bar{\gamma}) $ is
shown  in Fig.\ \ref{fig:DrudeP}.
Physically  $D ( \bar{\gamma} )$ is the dimensionless renormalization 
factor for the weight of the Drude peak, 
with $D = 1$ corresponding to an unrenormalized Drude peak.
The leading terms in the 
expansion of $D ( \bar{\gamma} )$ for small and large
$\bar{\gamma}$ are
 \begin{equation}
 D ( \bar{\gamma} ) \sim
 \left\{
 \begin{array}{ll}
 \frac{\sqrt{\pi}}{2} \bar{\gamma} &  \; , \; \bar{\gamma} \ll 1  \\
 1 & \; , \; \bar{\gamma} \gg 1
 \end{array}
 \right.
 \; .
 \label{eq:Drudeasym}
 \end{equation}
At the first sight the existence of a Drude peak
in our model is rather surprising because in
Sec.\ \ref{subsec:localization} we have found that the localization length
$\ell ( 0)$ at zero frequency is finite.
In fact, we believe that for Gaussian disorder
with moments given by Eqs.\ (\ref{eq:deltaav}) and (\ref{eq:deltacov})
the conductivity of the
one-dimensional FGM 
does not exhibit a Drude peak, because
the eigenstates at ${\omega} = 0$ should all be localized 
{\it{for a given realization of the disorder}} \cite{Lifshits88,Sadovskii91}.
On the other hand, for our choice   $\Delta ( x ) =A e^{i Q x}$ with
spatially constant but random $A$ and $Q$,  
the Green's function is not self-averging, so
that its spatial average 
is not identical with its disorder average.
As a consequence, 
there is a finite probability of finding
delocalized states at the Fermi energy:  for
$| \omega - \eta | > | A |$  the solutions of the
Schr\"{o}dinger equation are simply plane waves, whereas for 
$| \omega - \eta | < | A |$  there is a gap in the spectrum, 
and the Schr\"{o}dinger equation does not have any normalizable solutions.
Hence, depending on the realization of the disorder, the
system is either a perfect conductor or an insulator.
Because in Eq.\ (\ref{eq:Thouless}) we have defined the
inverse localization length in terms of the
{\it{disorder averaged}}  Green's function,
the value of
$\ell^{-1} ( \omega )$ is determined by those realizations
of the disorder where localized states at energy $\omega$ do not exist.
However, the probability of finding delocalized states
at the Fermi energy is finite, and 
can be expressed in terms of the function $P ( \bar{\gamma})$
defined in Eq.\ (\ref{eq:Pinfzero}),
 \begin{eqnarray}
 W_{\rm deloc} ( 0 )
 & = & \langle \Theta ( \eta^2 - | A |^2 ) \rangle
 \nonumber
 \\
 & = &
 1 -
 \frac{2}{\sqrt{\pi}} P ( \bar{\gamma} )
 \nonumber
 \\
 & \sim &
 \left\{
 \begin{array}{ll}
 \frac{2}{\sqrt{\pi}} \bar{\gamma} & , \; \bar{\gamma} \ll 1 \\
 1 & , \; \bar{\gamma} \gg 1
 \end{array}
 \right.
 \label{eq:Wres}
 \; .
 \end{eqnarray}
A graph if $W_{\rm deloc} ( 0 )$ is shown
as the dashed-dotted line in Fig.\ \ref{fig:DrudeP}.
Note that the qualitative behavior of
$W_{\rm deloc} ( 0 )$ is very similar to the weight
$D$ of the Drude peak.
The conductivity of quasi-one-dimensional Peierls systems {\it{below}}
the Peierls transition (for which $\langle \Delta ( x ) \rangle \neq 0$)
has been discussed in
Refs. \cite{Froehlich54,Lee74}. The authors pointed out that in this
case a gapless collective mode 
associated with fluctuations of the phase of the order parameter
generates a finite Drude peak. In our toy model, $\eta$ describes such
a gapless mode.

As discussed in Sec.\ \ref{sec:intro}, our model is also relevant
to describe higher-dimensional systems such as superconductors
within a quasiclassical approximation.  
In this case it is physically reasonable to expect that phase fluctuations of the superconducting
order parameter generate delocalized states
at the Fermi energy \cite{Emery95,Franz98}. Then we indeed
expect a finite Drude peak in the conductivity, which is 
broadened by disorder and becomes a sharp $\delta$-function
in the superconducting state.

\begin{figure}
\begin{center}
\psfrag{om}{$\bar{\omega}$}
\psfrag{C}{$\hspace{-5mm} C ( \bar{\gamma}, \bar{\omega})$}
\psfrag{0}{0}
\psfrag{0.5}{0.5}
\psfrag{1}{1}
\psfrag{0}{0}
\psfrag{1}{1}
\psfrag{2}{2}
\psfrag{3}{3}
\psfrag{4}{4}
\epsfig{file=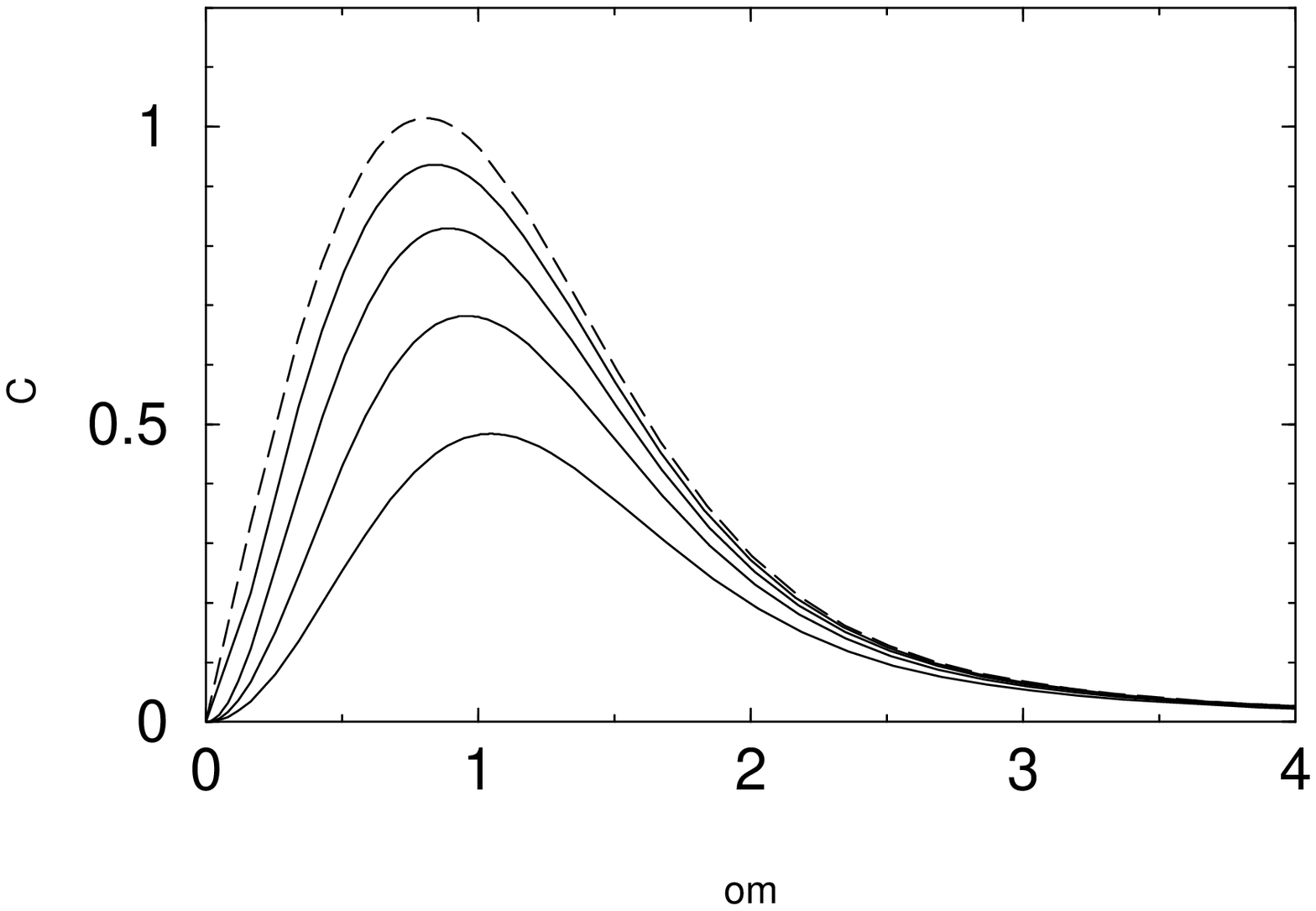,width=6cm}
\end{center}
\caption{Incoherent part 
$C ( \bar{\gamma} , \bar{\omega} )$
of the conductivity 
as a function of $\bar{\omega} = \omega / \Delta_s$,
see Eq.\ (\ref{eq:Cdef}). 
From top to bottom: 
$\bar{\gamma} = 0$ (dashed line) and  $\bar{\gamma} = 0.1, 0.25, 0.5, 1$.
} 
\label{fig:C}
\end{figure}

Let us now focus on the incoherent part of the
conductivity, which is described by the
dimensionless function $C ( \bar{\gamma} , \bar{\omega } )$
in Eq.\ (\ref{eq:ResigmaDC}).
A graph of this function is shown in Fig.\ \ref{fig:C}.
For large correlation lengths, i.e. $\bar{\gamma} \ll 1$ there are
three characteristic regimes 
where $C ( \bar{\gamma} , \bar{\omega } )$ can be
approximated by
 \begin{equation}
 C ( \bar{\gamma} , \bar{\omega} )
 \sim 
 \left\{
 \begin{array}{ll}
 \frac{4}{3}  \bar{\gamma}^{-1} \bar{\omega}^2 
 & \; , \; | \bar{\omega} | \ll \bar{\gamma} \\
 \frac{4\pi }{6} | \bar{\omega} |
 & \; , \; | \bar{\gamma} | \ll | \bar{\omega} | \ll 1 \\
 \frac{\pi }{2} | \bar{\omega} |^{-3}
 & \; , \; 1 \ll  | \bar{\omega} |  
 \end{array}
 \right.
 \; .
 \label{eq:Casym}
 \end{equation}
For $\bar{\gamma} \ll | \bar{\omega} |$ this agrees with 
the result of Ref. \cite{Sadovskii74}.
Note that for a one-band model with Gaussian white noise disorder the real
part of the conductivity is known to vanish for small
frequencies as $\omega^2 \ln^2 (1/ \omega)$ \cite{Mott67}. Thus, apart from the logarithmic
correction, the incoherent part of the conductivity of our simple model
shows the generic behavior of one-dimensional disordered electrons.
Note also that for small $\bar{\gamma}$ the relative weight of the
Drude peak is of the order of $\bar{\gamma}$, so that
the incoherent contribution dominates.

The white-noise limit is defined by letting $\Delta_s \xi \rightarrow
0$ while keeping $\Delta_s^2 \xi$ finite. In this case $D(\bar
\gamma)$ approaches unity.
In fact, in the white-noise limit the average conductivity is 
not modified by the disorder at all  because
the function $\Delta_s^{-1} C ( \bar{\gamma} , \bar{\omega} )$
vanishes if we let $\Delta_s \rightarrow \infty$. 

\section{Conclusions}
\label{sec:conclusions}

In this work we have introduced a simple exactly solvable toy model
which describes
the combined effects of phase and amplitude fluctuations
of an off-diagonal order parameter 
on the physical properties of an electronic system.
Although we have only discussed the one-dimensional version
of this model with linearized energy dispersion, 
the exact solubility of our model does not depend on these
features, so that our calculations can be generalized to
more realistic models of electrons in dimensions $d > 1$ with
non-linear energy dispersions.
In this case the fluctuating gap should be chosen
of the from
$\Delta ( {\bf{r}}) = A e^{ i {\bf{Q}} \cdot {\bf{r}}}$.
To satisfy $\langle \Delta ( {\bf{r}}) \rangle = 0$ and
$\langle \Delta ( {\bf{r}}) \Delta^{\ast} ( {\bf{r}}^{\prime}) 
\rangle = \Delta_s^2 e^{ - | {\bf{r}} - {\bf{r}}^{\prime}| / \xi}$,
the random variable $A$ should be distributed such that
Eqs.\ (\ref{eq:Afirstmom}) and (\ref{eq:Asecondmom}) are satisfied,
while the distribution ${\cal{P}}_{\bf{Q}}$
of the $d$-dimensional random-vector ${\bf{Q}}$ should be
 \begin{equation}
 {\cal{P}}_{\bf{Q}}=
 \frac{ 1}{( 2 \pi)^d}
 \int d {\bf{r}} e^{ - i {\bf{Q}} \cdot {\bf{r}}}
 e^{ - | {\bf{r}}| / \xi }
 \; .
 \label{eq:Probarb}
 \end{equation}
For $d=1$ this reduces to  Eq.\ (\ref{eq:peps}),
but in $d>1$ Eq.\ (\ref{eq:Probarb}) is not a Lorentzian.
 
In one dimension our model
describes the disordered phase
of Peierls and spin-Peierls chains. 
We have presented explicit results
for the density of states, the localization length, the 
single-particle spectral function, and the real part of the
conductivity.
Let us emphasize three points:

(a) The mean localization length of our toy model, which
we have defined via the Thouless formula (\ref{eq:Thouless}),
is an excellent approximation to the
mean localization length of the FGM with Gaussian disorder.
Although the respective density of states agree quite well on a
qualitative level, deviations become substantial for large correlation
lengths, leading to a different scaling behavior as a function of
$\xi$.

(b) The interplay between phase and amplitude fluctuations
gives rise to a weak logarithmic singularity in the single-particle
spectral function of our model. Whether this singularity is just an
artifact of our toy model or not remains an open question.

(c) The conductivity of our model exhibits not only a
pseudogap below the energy scale $\Delta_s$ but also 
a Drude peak at $\omega = 0$ with a weight that vanishes
as $1/ \xi$ for $\xi \rightarrow \infty$.
While the qualitative picture of the continuous part should be generic
for more realistic 
one-dimensional disordered systems (up to 
logarithmic corrections for small frequencies \cite{Mott67}), the
Drude peak in our model is due to the existence of delocalized states
at the Fermi energy 
which are created by phase fluctuations. However, in a strictly
one-dimensional disordered system, the disorder should lead to the localization
of all eigenstates, resulting in a vanishing zero temperature dc
conductivity \cite{Mott67}. On the other hand, even very weak 
three-dimensional interactions can lead to a phase transition leading
to long-range order and a finite Drude peak as found in our model.
We expect that 
forward scattering by  disorder (which we have ignored in our
calculation)
will broaden the Drude peak \cite{Kopietz99}.
Experimentally, peak structures in the far infrared well below the 
pseudogap regime have been
observed in the optical conductivity
of several quasi one-dimensional Peierls systems above
the Peierls transition \cite{Gorshunov94}.

Our model also  describes superconducting
fluctuations in $d>1 $  within  a semiclassical approximation.
Recall that our
Eq.\ (\ref{eq:Andreev}) for the Green's function in $d=1$ 
is formally equivalent to the
Andreev equation for the semiclassical wave-function of a 
superconductor. The latter can be obtained
from the more general Gorkov equation (\ref{eq:Gorkov})
in the limit of a slowly varying order parameter.
To calculate
physical observables, the solutions of the
Andreev equations should be averaged over the
classical trajectories of the electrons \cite{Andreev64}, which we have
not done in this work. Therefore we cannot make any quantitative
comparisons with experimental data for high-temperature
superconductors.
However, some qualitative features of our results seem
to agree with experiments. In particular, in our model
the pseudogap in  the conductivity
coexists with a small Drude peak. 
Such a behavior has been seen experimentally
in the normal state
of high-temperature superconductors \cite{Lupi00}.
In our model the Drude peak  
is a direct consequence of the fluctuating phase of the
superconducting order parameter. Without phase
fluctuations all charge carriers at the Fermi energy are localized and
there is no Drude peak.
In this respect our model
describes a bad metal in the sense defined by Emery and Kivelson
\cite{Emery95}. 

\begin{acknowledgement}

This work was financially supported by the
DFG (Grants No. Ko 1442/3-1 and Ko 1442/4-1).
\end{acknowledgement}

%

%
%
%
%
%
%
%
%
%
%
%
%
%
%
%
%
%
%
%
%
%
%
%

\begin{thebibliography}{99}
\bibitem{Sadovskii79}
M. V. Sadovskii, Zh. Eksp. Teor. Fiz. {\bf{77}},
2070 (1979) [Sov. Phys. JETP {\bf{50}}, 989 (1979)].
Note that the parameter $\Gamma $ defind
by Sadovskii can be identified with our
$2 \bar{\gamma} = v_F / ( \Delta_s \xi )$.
\end{thebibliography}
\end{document}